\documentclass[
    ,final            % use final for the camera ready runs
  ]
  {aipproc}

\layoutstyle{6x9}

\newcommand{\shat}{\hat{s}^2}

\newcommand{\rwhat}{\Delta\hat{r}_W}
\newcommand{\rzhat}{\Delta\hat{r}_Z}
\newcommand{\dgama}{\hat\Delta_\gamma}

\newcommand{\etal}{{\em et al.}}
\newcommand{\ibid}{{\em ibid.}}
\newcommand{\be}{\begin{equation}}
\newcommand{\ee}{\end{equation}}
\newcommand{\bea}{\begin{eqnarray}}
\newcommand{\eea}{\end{eqnarray}}
\newcommand{\ba}{\begin{array}}
\newcommand{\ea}{\end{array}}

\begin{document}

\title{Electroweak Standard Model and Precision Tests}

\author{Jens Erler\footnote{Talk given at the X Mexican
School of Particles and Fields, Playa del Carmen, M\'exico, 2002.}}{
address={Instituto de F\'\i sica, Universidad Nacional Aut\'onoma de M\'exico,
           01000 M\'exico D.F., M\'exico}}

\begin{abstract}
I give an introduction and overview of recent developments in high precision 
tests of the Standard Model.  This includes a summary of $Z$-pole measurements,
a brief account of the NuTeV result on neutrino-nucleon scattering, the
anomalous magnetic moment of the muon, and implications for the Higgs boson 
mass.
\end{abstract}

\maketitle

%%%%%%%%%%%%%%%%%%%%%%%%%%%%%%%%%%%%%%%%%%%%
%% MAINMATTER
%%%%%%%%%%%%%%%%%%%%%%%%%%%%%%%%%%%%%%%%%%%%

\section{Introduction}
The most fundamental observable related to the weak interaction is the muon
lifetime, $\tau_\mu$.  With the electromagnetic two-loop contribution obtained 
in Ref.~\cite{vanRitbergen:1998yd}, $\tau_\mu$ can be used unambiguously 
to extract the Fermi constant, $G_F=1.16637(1)\times 10^{-5}\mbox{ GeV}^{-2}$,
where the uncertainty is completely dominated by experiment.  
With the electromagnetic fine structure constant, $\alpha$, as an additional 
input, we can then obtain the quantity,
\be
   A^2 =\frac{\pi\alpha}{\sqrt{2} G_F}=(37.2805 \pm 0.0003 \hbox{ GeV})^2,
\ee
and use it to write down relations between the intermediate gauge boson masses,
$M_{W,Z}$, and the weak mixing angle; for example, in the $\overline{\rm MS}$ 
scheme one has~\cite{Degrassi:1990tu},
\be 
\sin^2\hat\theta_W \equiv \shat = \frac{A^2}{M_W^2 (1 - \rwhat)}, \hspace{50pt}
                     \shat (1 - \shat) = \frac{A^2}{M_Z^2 (1 - \rzhat)},
\label{rzwhat}
\ee
where $\rwhat$ and $\rzhat$ are electroweak radiative correction parameters.
Most of the $Z$-pole asymmetries discussed in the next section are basically
measurements of $\sin^2\theta^{\rm eff}_e = \hat\kappa_e\shat$, where 
$\hat\kappa_f$ denotes a flavor dependent form factor which for $f = e$ is 
numerically very close to unity with little sensitivity to the Standard Model 
(SM) input parameters. Since furthermore $M_Z$ is known to great accuracy, 
the second Eq.~(\ref{rzwhat}) implies that the $Z$-pole asymmetries effectively
determine,
\be
    \rzhat = {\alpha\over\pi} \dgama + F_1(m_t^2,M_H,\dots).
\label{rzhat}
\ee
Asymptotically for large top quark masses, $m_t$, the function, $F_1$, grows 
like $m_t^2$.  This is because the large mass hierarchy, $m_t \gg m_b$, breaks 
isospin symmetry in $W$ boson self-energy diagrams with a bottom and a top 
quark in the loop.  This effect has been absorbed into $G_F$, but now reappears
in $\rzhat$ when $M_Z$ is computed in terms of it.  For the same reason, there 
is an $m_t^2$ effect in the low-energy $\rho$-parameter~\cite{Veltman:1977kh}, 
which is defined as the ratio of neutral-to-charged weak current interaction 
strengths.  The first Eq.~(\ref{rzwhat}) shows that a determination of 
the $W$ boson mass can then be used to measure
\be
    \rwhat = {\alpha\over\pi} \dgama + F_2(\ln \ m_t,M_H,\dots),
\label{rwhat}
\ee
where indeed $F_2$ has a milder $m_t$ dependence.  $F_1$ and $F_2$ are 
complicated functions of the Higgs boson mass, $M_H$, which are asymptotically 
logarithmic.  Eqs.~(\ref{rzhat}) and (\ref{rwhat}) also show that $M_H$ can be
extracted from the precision data only when the quantity,
\be
   \dgama = 4\pi^2 \hat\Pi_{\gamma\gamma}^{(f)} + \mbox{ bosonic terms},
\ee
which characterizes the renormalization group (RG) evolution (``running'') of 
$\alpha$,
\be
   \hat\alpha (M_Z) = {\alpha\over 1 - {\alpha\over\pi} \dgama},
\ee
is known accurately.  While it can be computed rigorously for leptons, there is
a problem for quarks (hadrons).   This is best seen by noting that the one-loop
fermion contribution to the photon self-energy, $\hat\Pi_{\gamma\gamma}^{(f)}$,
is proportional to $\ln M_Z^2/m_f^2$, and it is not clear what mass definition
is to be used here for quarks.  This is a question of loop corrections 
proportional to powers of the QCD coupling, $\alpha_s$, and can indeed be dealt
with perturbatively for charm and bottom quarks~\cite{Erler:1998sy}.  
For the light quarks, however, perturbation theory breaks down and one needs
a different strategy:  one uses analyticity and the optical theorem which in 
essence delivers $\hat\Pi_{\gamma\gamma}^{(f)}$ from its imaginary part and 
thus from (a weighted integral over) the cross-section 
$\sigma (e^+ e^- \rightarrow {\rm hadrons})$ for which experimental data are 
available.  Incidentally, a similar strategy is used to estimate the hadronic 
two- and three-loop contributions to the muon anomalous magnetic moment, 
$g_\mu-2$, which amounts to an integral over the same data, but with 
a different weight.  Notice that the uncertainty in the cross-section data 
induces correlations between $\dgama$, $M_H$, $g_\mu-2$, and (currently of less
importance) the running of the weak mixing angle relevant for weak neutral 
current precision observables at low energies (such as in atomic parity 
violation) which is also subject to this kind of treatment and the same data.

By assuming isospin symmetry and correcting for kinematics, isospin violating
effects~\cite{Cirigliano:2001er}, electroweak radiative 
corrections~\cite{Marciano:pd,Braaten:1990ef,Erler:2002mv}, etc., one can use 
the invariant mass spectrum in hadronic $\tau$ decays to obtain additional 
information~\cite{Davier:2002dy}.  Kinematic suppression limits this method 
mainly to two pion (and to a lesser extent four pion) final states. Hadronic 
$\tau$ decay data are of particular relevance to $g_\mu-2$ (see below).

\section{$Z$-pole observables}
\label{zpole}

\begin{table}
\begin{tabular}{llcccr}
\hline
Quantity & & Group(s) & Value & Standard Model & pull\\
\hline
$M_Z$ & [GeV] & LEP &$ 91.1876 \pm 0.0021 $&$ 91.1874 \pm 0.0021 $&$ 0.1$\\
$\Gamma_Z$ & [GeV] & LEP &$ 2.4952 \pm 0.0023 $&$  2.4972 \pm 0.0011 $&$-0.9$\\
$\Gamma({\rm inv})$ & [MeV] & LEP &$ 499.0\pm 1.5   $& $501.74\pm 0.15$  &---\\
$\sigma_{\rm had}$ & [nb] & LEP & $41.541\pm 0.037$ & $41.470\pm 0.010$&$1.9$\\
$R_e$    & & LEP & $20.804 \pm 0.050$ & $20.753 \pm 0.012$ & $ 1.0$\\
$R_\mu$  & & LEP & $20.785 \pm 0.033$ & $20.753 \pm 0.012$ & $ 1.0$\\
$R_\tau$ & & LEP & $20.764 \pm 0.045$ & $20.799 \pm 0.012$ & $-0.8$\\
$A_{FB} (e)$ & & LEP &   $0.0145 \pm 0.0025$ & $0.01639 \pm 0.00026$ & $-0.8$\\
$A_{FB} (\mu)$ & & LEP & $0.0169 \pm 0.0013$ & $                   $ & $ 0.4$\\
$A_{FB} (\tau)$ & & LEP& $0.0188 \pm 0.0017$ & $             $ & $ 1.4$\\
\hline
$R_b$ & & LEP + SLD & $0.21644 \pm 0.00065$ & $0.21572 \pm 0.00015$ & $ 1.1$\\
$R_c$ & & LEP + SLD & $ 0.1718 \pm 0.0031 $ & $0.17231 \pm 0.00006$ & $-0.2$\\
$A_{FB} (b)$ & & LEP & $0.0995 \pm 0.0017$ & $0.1036 \pm 0.0008$ & $-2.4$\\
$A_{FB} (c)$ & & LEP & $0.0713 \pm 0.0036$ & $0.0741 \pm 0.0007$ & $-0.8$\\
$A_b$ & & SLD & $0.922 \pm 0.020$ & $0.93477 \pm 0.00012$ & $-0.6$\\
$A_c$ & & SLD & $0.670 \pm 0.026$ & $0.6681  \pm 0.0005 $ & $ 0.1$\\
\hline
$A_{LR}$ (hadrons) \hspace*{-20pt} & & SLD & $0.15138\pm 0.00216$ & $0.1478\pm 0.0012$&$ 1.6$\\
$A_{LR}$ (leptons) \hspace*{-20pt} & & SLD & $0.1544 \pm 0.0060 $ &                   &$ 1.1$\\
$A_\mu$            & & SLD & $0.142  \pm 0.015  $ &                   &$-0.4$\\
$A_\tau$           & & SLD & $0.136  \pm 0.015  $ &                   &$-0.8$\\
$A_\tau ({\cal P}_\tau)$ & & LEP & $0.1439 \pm 0.0043$ &              &$-0.9$\\
$A_e ({\cal P}_\tau)$    & & LEP & $0.1498 \pm 0.0049$ &              &$ 0.4$\\
$Q_{FB}$ & & LEP & $0.0403 \pm 0.0026$ & $0.0424 \pm 0.0003 $ & $-0.8$\\
\hline
\end{tabular}
\caption{Results from $Z$-pole precision measurements compared to the SM 
predictions obtained from a global analysis of high and low energy experiments.
The deviations from the predictions (in terms of the pulls) are also shown.}
\label{tab:zpole}
\end{table}     

The first part of Table~\ref{tab:zpole} shows the $Z$ line shape and leptonic 
forward-backward (FB) cross section asymmetry, $A_{FB}(\ell)$, measurements 
from LEP~1~\cite{Abbaneo:2001ix}.  They include the total $Z$ decay width, 
$\Gamma_Z$, the hadronic peak cross section, $\sigma_{\rm had}$, and for each 
lepton flavor the ratio of hadronic to leptonic partial $Z$ widths, $R_\ell$.  
The invisible $Z$ partial width, $\Gamma({\rm inv})$, is derived from 
$\Gamma_Z$, $\sigma_{\rm had}$, and the $R_\ell$, and is not independent. It is
smaller than the SM prediction by almost 2~$\sigma$, which can be traced to 
$\sigma_{\rm had}$ which deviates by a similar amount.  Conversely, one can 
use the data to determine the number of standard neutrinos, 
$N_\nu = 2.986 \pm 0.007$, again showing a 2~$\sigma$ deviation from 
the expectation, $N_\nu=3$.

The second part of Table~\ref{tab:zpole} shows the results from $Z$ decays into
heavy flavors~\cite{Abbaneo:2001ix}.  For bottom and charm quarks the partial 
$Z$ width normalized to the hadronic partial width, $R_q$, is shown, as well as
the FB-asymmetry, $A_{FB}(q)$, and $A_q$ which is proportional to the combined 
left-right (LR) forward-backward asymmetry, $A_{FB}^{LR}(q)$.  The latter is 
equivalent to $\sin^2\theta^{\rm eff}_q$.  $A_{FB}(q)$ is proportional to 
$A_e A_q$ and primarily sensitive to $\sin^2\theta^{\rm eff}_e$ with 
$A_{FB}(b)$ providing one of its best determinations. It shows a 2.4~$\sigma$ 
deviation, and favors larger values of $M_H$. It is tempting to suggest new
physics effects in the factor $A_b$ to reconcile this deviation and 
the disagreement with $A_{LR}$ discussed below. However, one would need 
a $(19 \pm 7)$\% radiative correction to $\kappa_b$ while typical electroweak 
radiative corrections are of ${\cal O} (1\%)$ or smaller. New physics entering
at tree level is generally not resonating and/or strongly constrained by other
processes. At any rate, $R_b$ is in reasonable agreement with the SM and some 
tuning of parameters would be required.

The last part of Table~\ref{tab:zpole} shows further measurements with 
sensitivity to $\sin^2\theta^{\rm eff}_\ell$.  The LR cross section asymmetry,
 $A_{LR} = A_e$, from the SLD Collaboration for hadronic~\cite{Abe:2000dq} and 
leptonic final states~\cite{Abe:2000hk} show a combined deviation of 
1.9~$\sigma$ from the SM prediction. In contrast to $A_{FB}(b)$, it favors 
small values of $M_H$, which are excluded by the direct searches at 
LEP~2~\cite{Holzner:2002ft}, 
\be
   M_H \geq 114.4 \mbox{ GeV } (95\% \mbox{ CL}).
\label{lep2limit}
\ee
The other determinations are from $A_{FB}^{LR}(\mu)$ and 
$A_{FB}^{LR}(\tau)$~\cite{Abe:2000hk}, from the final state $\tau$ 
polarization, ${\cal P}_\tau$, and its angular 
dependence~\cite{Abbaneo:2001ix}, as well as the hadronic charge 
asymmetry~\cite{Abbaneo:2001ix}, which is a weighted sum over $A_{FB}(q)$.

\section{Other observables}
\label{nonzpole}

\begin{table}
\begin{tabular}{lcccr}
\hline Quantity & Group(s) & Value & Standard Model & pull \\ 
\hline
$m_t$\hspace{8pt}[GeV]&Tevatron &$ 174.3    \pm 5.1               $&$ 174.4    \pm 4.4    $&$ 0.0$ \\
$M_W$ [GeV]    &      LEP       &$  80.447  \pm 0.042             $&$  80.391  \pm 0.019  $&$ 1.3$ \\
$M_W$ [GeV]    & Tevatron + UA2 &$  80.454  \pm 0.059             $&$                     $&$ 1.1$ \\
\hline
$g_L^2$        &     NuTeV      &$   0.30005\pm 0.00137           $&$   0.30396\pm 0.00023$& $-2.9$ \\
$g_R^2$        &     NuTeV      &$   0.03076\pm 0.00110           $&$   0.03005\pm 0.00004$&$ 0.6$ \\
$R^\nu$        &     CCFR       &$   0.5820 \pm 0.0027 \pm 0.0031 $&$   0.5833 \pm 0.0004 $&$-0.3$ \\
$R^\nu$        &     CDHS       &$   0.3096 \pm 0.0033 \pm 0.0028 $&$   0.3092 \pm 0.0002 $&$ 0.1$ \\
$R^\nu$        &     CHARM      &$   0.3021 \pm 0.0031 \pm 0.0026 $&$                     $& $-1.7$ \\
$R^{\bar\nu}$  &     CDHS       &$   0.384  \pm 0.016  \pm 0.007  $&$   0.3862 \pm 0.0002 $&$-0.1$ \\
$R^{\bar\nu}$  &     CHARM      &$   0.403  \pm 0.014  \pm 0.007  $&$                     $&$ 1.0$ \\
$R^{\bar\nu}$  &     CDHS 1979  &$   0.365  \pm 0.015  \pm 0.007  $&$   0.3817 \pm 0.0002 $&$-1.0$ \\
\hline
$g_V^{\nu e}$  &     CHARM II   &$  -0.035  \pm 0.017             $&$  -0.0398 \pm 0.0003 $&  ---  \\
$g_V^{\nu e}$  &      all       &$  -0.041  \pm 0.015             $&$                     $&$-0.1$ \\
$g_A^{\nu e}$  &     CHARM II   &$  -0.503  \pm 0.017             $&$  -0.5065 \pm 0.0001 $&  ---  \\
$g_A^{\nu e}$  &      all       &$  -0.507  \pm 0.014             $&$                     $&$ 0.0$ \\
\hline
$Q_W({\rm Cs})$&     Boulder    &$ -72.69   \pm 0.44              $&$ -73.10   \pm 0.04   $&$ 0.8$ \\
$Q_W({\rm Tl})$&Oxford + Seattle&$-116.6    \pm 3.7               $&$-116.7    \pm 0.1    $&$ 0.0$ \\
\hline
%$\Gamma (b\rightarrow s\gamma)/\Gamma_{SL}$ & BaBar + Belle + CLEO &$3.48^{+0.65}_{-0.54} \times 10^{-3}$ & $3.20 \pm 0.09 \times 10^{-3}$&$ 0.5$ \\
%\hline
%$\tau_\tau$ [fs] & direct + ${\cal B}_e + {\cal B}_\mu$ &$ 290.96 \pm 0.59 \pm 5.66 $&$ 291.90 \pm 1.81 $&$-0.4$ \\
%$10^4$ $\Delta\alpha^{(3)}_{\rm had}$ & $e^+e^-$ data + $\tau$ decays &$ 56.53 \pm 0.83 \pm 0.64 $&$ 57.52 \pm 1.31 $&$-0.9$ \\
$10^9$ $(a_\mu - {\alpha\over 2\pi})$ & BNL + CERN &$ 4510.64 \pm 0.79 \pm 0.51 $&$ 4508.28 \pm 0.33 $& $2.5$ \\
\hline
\end{tabular}
\caption{Precision observables away from the $Z$-pole. The first error for
the measurement values is experimental and (where applicable) the second refers
to theory or model uncertainties.}
\label{tab:lowenergy}
\end{table}

\begin{figure}
  \includegraphics[height=.5\textheight]{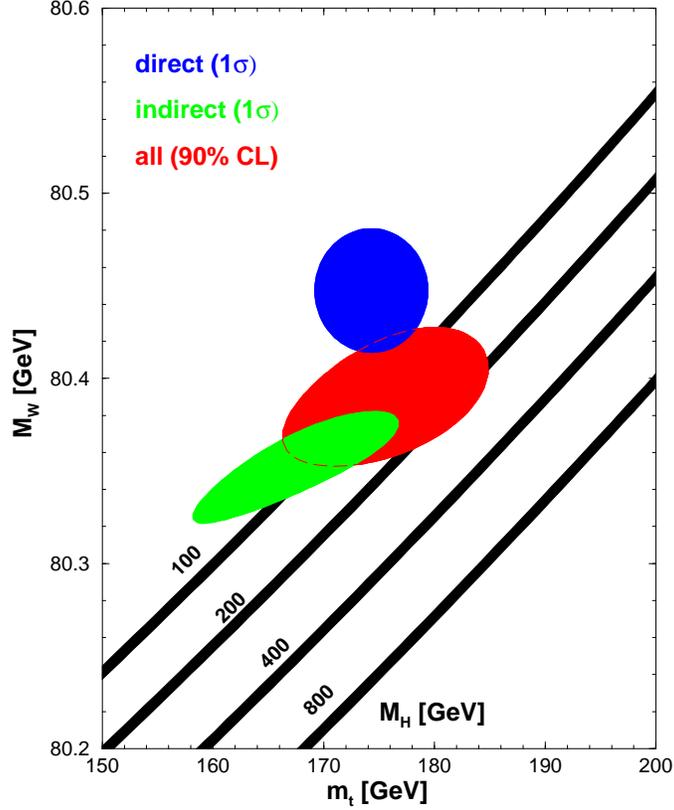}
  \caption{One-standard-deviation (39.35\% CL) regions in the $M_W$-$m_t$ plane
  for the direct and indirect data.  The combined 90\% CL contour 
  ($\Delta\chi^2 = 4.605$) is also shown.  The widths of the $M_H$ bands 
  represent the theoretical uncertainty in the SM prediction 
  ($\alpha_s (M_Z) = 0.120$).
\label{mwmt}}
\end{figure}

The first part of Table~\ref{tab:lowenergy} shows the (direct) $m_t$ 
measurement from the Tevatron~\cite{Abbott:1998dc,Abe:1998bf}, as well as $M_W$
from LEP~2~\cite{Abbaneo:2001ix} and $p\bar{p}$ 
collisions~\cite{Abbott:1999ns,Affolder:2000bp}.  The combined
$M_W$ is 1.8~$\sigma$ higher than the SM expectation.  Just as $A_{LR}$ it 
favors smaller values of $M_H$.  We compare these mass measurements with all 
other (indirect) data, and the SM prediction for various values of $M_H$ in 
Figure~\ref{mwmt}.  The bottom and charm quark masses, $m_b$ and $m_c$, which 
enter the SM predictions of numerous observables (for example through $\dgama$)
are constrained using a set of inclusive QCD sum rules~\cite{Erler:2002bu} and 
are recalculated in each call within the fits as functions of $\alpha_s$ and 
other global fit parameters.

\begin{figure}
  \includegraphics[height=.5\textheight]{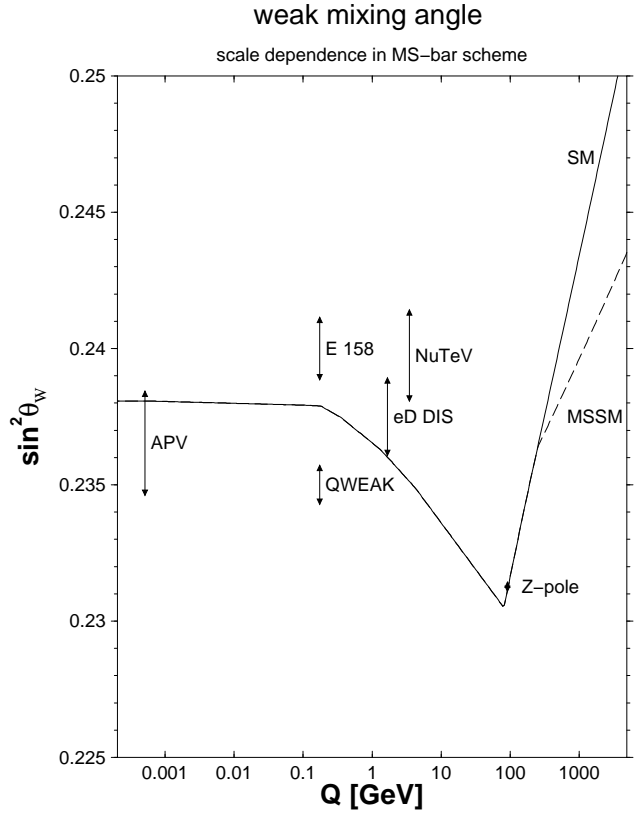}
  \caption{Scale dependence of the weak mixing angle in the SM and the minimal
supersymmetric SM.  The error bars for E158, QWEAK, and eD DIS are projections
(the vertical location is arbitrary).
  \label{sin2theta}}
\end{figure}

The second part of Table~\ref{tab:lowenergy} shows results of neutrino-nucleon
deep inelastic scattering experiments from CERN and FNAL.  The ratio of 
neutral-to-charged current $\nu_\mu$-cross sections,
$R_\nu = \sigma_{\nu N}^{\rm NC}/\sigma_{\nu N}^{\rm CC}$, is more sensitive to
the weak mixing angle than the analogous $\bar\nu_\mu$-ratio, $R_{\bar\nu}$.  
Both are sensitive to charm threshold effects which introduce the dominant 
theoretical uncertainty. This uncertainty largely cancels in 
the ratio~\cite{Paschos:1972kj},
\be
   R^- = \frac{\sigma_{\nu N}^{\rm NC} - \sigma_{\bar\nu N}^{\rm NC}}
              {\sigma_{\nu N}^{\rm CC} - \sigma_{\bar\nu N}^{\rm CC}}
       = g_L^2 - g_R^2,
\ee
which was used by the NuTeV Collaboration~\cite{Zeller:2001hh} (who had 
a clean $\bar\nu_\mu$-beam at their disposal) to measure the weak mixing angle 
precisely off the $Z$-pole. In the presence of new physics, however, which will
in general affect $\nu_\mu$ and $\bar\nu_\mu$ cross sections differently, one 
should rather monitor $R_\nu$ and $R_{\bar\nu}$ independently, or equivalently,
the effective four-Fermi $\nu_\mu$-quark couplings, 
\be
   g_L^2 = {1\over 2} - \sin^2\theta_W + {5\over9} \sin^4\theta_W,\hspace{50pt}
   g_R^2 = {5\over9} \sin^4\theta_W,
\ee
which are shown in the Table.  One sees that the quoted~\cite{Zeller:2001hh} 
3~$\sigma$ deviation in the weak mixing angle can be traced to a 2.9~$\sigma$
deviation in $g_L^2$. 
The weak charge of Cs~\cite{Bennett:1999pd}, 
$Q_W ({\rm Cs})$, is currently the only other precise determination of 
$\sin^2\theta_W$ off the $Z$-pole, while the results of $\nu e$ scattering in
the third part of Table~\ref{tab:lowenergy} and of $Q_W ({\rm Tl})$ are less 
precise.  The extraction of weak charges from atomic parity violation is 
complicated by atomic theory~\cite{Kuchiev:2002fg} uncertainties and 
$Q_W ({\rm Cs})$ was deviating from the SM prediction in the past.  Thus, it is
important to have complementary determinations of $\sin^2\theta_W$ which (as 
indicated in Figure~\ref{sin2theta}) will come from polarized M{\o}ller 
scattering by the E158 Collaboration at SLAC~\cite{Carr:1997fu} and elastic 
$\vec{e}p$ scattering by the QWEAK Collaboration at JLAB~\cite{Armstrong:2001}
which will determine the weak charges of the electron and the proton, 
respectively.  A new deep inelastic $eD$ scattering experiment has also been 
suggested~\cite{Reimer:2002}.

The last entry in Table~\ref{tab:lowenergy} refers to 
$g_\mu - 2$~\cite{Bennett:2002jb} which shows a deviation of about 2.5~$\sigma$
from the SM prediction.  This could be interpreted as a hint at supersymmetry 
or many other types of physics beyond the SM~\cite{Czarnecki:2001pv}.  However,
as discussed in the introduction, there are complications from the two-loop 
hadronic vacuum polarization.  In fact, data based on 
$\sigma (e^+ e^- \rightarrow {\rm hadrons})$ suggest even a 3~$\sigma$ 
deviation~\cite{Davier:2002dy,Hagiwara:2002ma}, while data based on $\tau$ 
decays suggest agreement with the SM within about one standard 
deviation~\cite{Erler:2002mv}. Moreover, the three-loop hadronic light-by-light
contribution can be modeled~\cite{Knecht:2001qg}, but methods based on first 
principles such as chiral perturbation theory are not precise enough to confirm
the model estimates~\cite{Ramsey-Musolf:2002cy}.

\begin{figure}
  \includegraphics[height=.5\textheight]{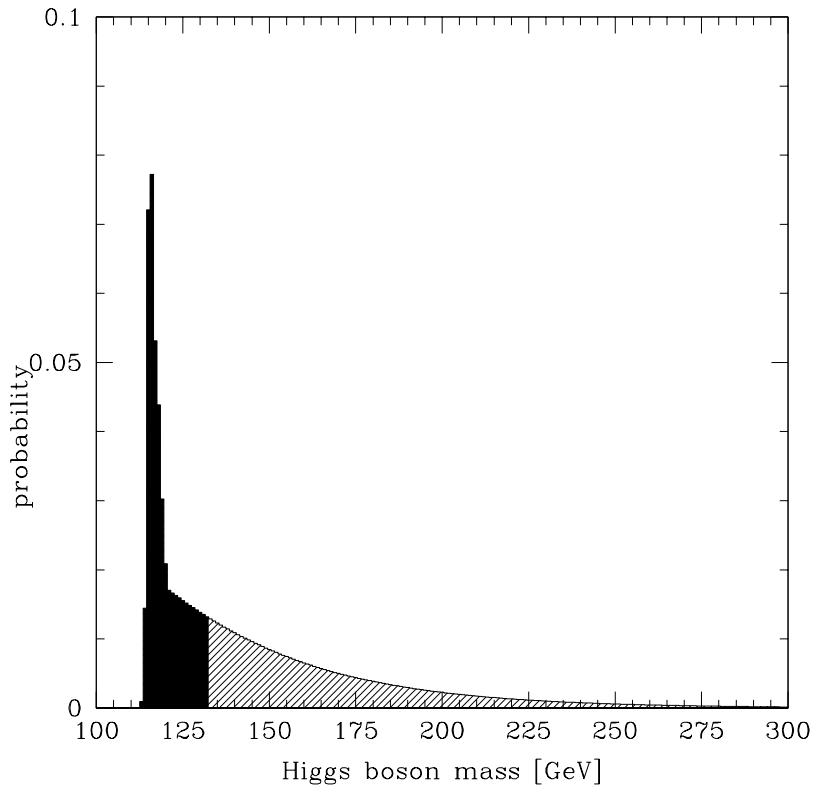}
  \caption{Probability density for $M_H$ obtained by combining precision data
  with direct search results at LEP~2.  The peak is due to the candidate Higgs
  events observed at LEP (updated from Ref.~\cite{Erler:2000cr}).
  \label{higgspdf}}
\end{figure}

\section{Conclusions}
\label{conclusions}
Despite of the various deviations described above, it must be stressed that 
the overall agreement between the data and the SM is reasonable. The $\chi^2$
per degree of freedom of the global fit is 49.1/40 where the probability for
a larger $\chi^2$ is 15\%.  

The global fit to all precision data currently favors values for the Higgs
boson mass\footnote{See Ref.~\cite{Frank:2002qa} and the talk of W. Hollik
at this meeting for a discussion of Higgs boson masses in the minimal
supersymmetric SM.},
\be
M_H = 86^{+49}_{-32}\mbox{ GeV}, \hspace{50pt} 
(M_H = 93^{+52}_{-35}\hbox{ GeV}),
\ee
where the central value is slightly below the lower LEP~2 limit in 
Eq.~(\ref{lep2limit}).  The result in parentheses refers to the case where
the $g_\mu - 2$ result is excluded from the fit.  The two results differ not
because the SM prediction for $g_\mu - 2$ has a strong $M_H$ dependence, but 
rather due to the correlation described in the Introduction.  If one includes 
the Higgs search information from LEP~2, one obtains 
the probability density shown in Figure~\ref{higgspdf}.  

\noindent
The global fit result,
\be 
   m_t = 174.2 \pm 4.4 \hbox{ GeV}, \hspace{50pt} 
  (m_t = 174.0^{+9.9}_{-7.4} \hbox{ GeV}),
\ee
is dominated by the Tevatron measurement, $m_t = 174.3 \pm 5.1$~GeV.  One can
exclude this from the fit and determine $m_t$ from the (indirect) precision 
data alone.  This is the result in parentheses, which is seen in spectacular 
agreement with the Tevatron value.

%%%%%%%%%%%%%%%%%%%%%%%%%%%%%%%%%%%%%%%%%%%%%%%%
%% BACKMATTER
%%%%%%%%%%%%%%%%%%%%%%%%%%%%%%%%%%%%%%%%%%%%%%%%

\begin{theacknowledgments}
It is a pleasure to congratulate Augusto Garc\'\i a and Arnulfo Zepeda to their
60th birthday, and to thank the organizers for inviting me to a very enjoyable 
meeting.
\end{theacknowledgments}

\end{document}